\documentclass[onecolumn,manuscript,showpacs,amsmath,apsrev4-1]{revtex4-1}
\usepackage{}
\usepackage{txfonts}
\usepackage{pifont}
\usepackage{amssymb}

\usepackage{dcolumn}
\usepackage{amsmath}
\usepackage[dvips]{epsfig}

\makeatletter
\def\btt#1{\texttt{\@backslashchar#1}}%
\DeclareRobustCommand\bblash{\btt{\@backslashchar}}%
\makeatother

\begin{document}
\title{A Three-Dimensional Tight-Binding Model and
Magnetic Instability of KFe$_{2}$Se$_{2}$}
\author{Da-Yong Liu$^{1}$, Ya-Min Quan$^{1}$, Zhi Zeng$^{1}$, Liang-Jian Zou$^{1,
        \footnote{Correspondence author, Electronic mail:
           zou@theory.issp.ac.cn}}$
%           , and Hai-Qing Lin$^{3}$
}
\affiliation{ \it $^1$ Key Laboratory of Materials Physics,
Institute of Solid State Physics,
Chinese Academy of Sciences, P. O. Box 1129, Hefei 230031,
People's Republic of China
%\\
%\it $^2$ College of Physics Science and Technology,
%China University of Petroleum,
%DongYing 257061, People's Republic of China
%\\
%\it $^3$ Department of Physics, Chinese University of Hong Kong,
%Shatin, New Territory, Hong Kong, China
\\}
\date{Mar 20, 2011}

\begin{abstract}

For a newly discovered iron-based high T$_{c}$ superconducting
parent material KFe$_{2}$Se$_{2}$,
we present an effective three-dimensional five-orbital
tight-binding model by fitting the band structures.
%obtained within density functional theory.
%
The three t$_{2g}$-symmetry orbitals of the five Fe 3d orbitals mainly
contribute to the electron-like Fermi surface, in agreement with
recent angle-resolved photoemission
spectroscopy experiments.
To understand the groundstate magnetic structure, the two- and
three-dimensional dynamical spin susceptibilities
within the random phase approximation are investigated.
It obviously shows a sharp peak at wave
vector $\mathbf{Q}$ $\thicksim$ ($\pi$, $\pi$), indicating the magnetic
instability of {\it N$\acute{e}$el}-type antiferromagnetic rather
than ($\pi$/2, $\pi$/2)-type antiferromagnetic ordering.
While along $\emph{c}$ axis, it exhibits a ferromagnetic coupling
between the nearest neighboring FeSe layers.

\end{abstract}

\maketitle

\section{INTRODUCTION}

% superconductivity Tc ¡¤
Since the discovery of the iron-based superconductor has been reported
\cite{JACS130-3296}, a series of superconducting compounds were
found, such as 1111 phase ({\it e.g.} LaOFeAs \cite{JACS130-3296}),
122 phase ({\it e.g.} BaFe$_{2}$As$_{2}$ \cite{PRL101-257003}), 111
phase ({\it e.g.} NaFeAs \cite{PRL102-227004}), and 11 phase ({\it
e.g.} FeSe \cite{PNAS105-14262}, FeTe \cite{PRL102-247001}).
Recently a new superconductor member KFe$_{2}$Se$_{2}$ with T$_{c}$
above 30 K (K$_{0.8}$Fe$_{2}$Se$_{2}$) \cite{PRB82-180520} has been
reported, and attracted considerable interests for its unique
insulator and large magnetic moment properties, quite different from
other iron-based superconducting materials for its parent material
KFe$_{2}$Se$_{2}$ is iso-structural to BaFe$_{2}$As$_{2}$, but
chemically close to FeSe.

% structure
The space group of KFe$_{2}$Se$_{2}$ at room temperature is I4/mmm
(No. 139) \cite{PRB82-180520}. Its crystal structure is
composed of edge-sharing FeSe$_{4}$
tetrahedra separated by K cations, as shown schematically in Fig. 1.
%
%-------------------------------------------------------------------------
\begin{figure}[htbp]\centering
\includegraphics[angle=0, width=0.3 \columnwidth]{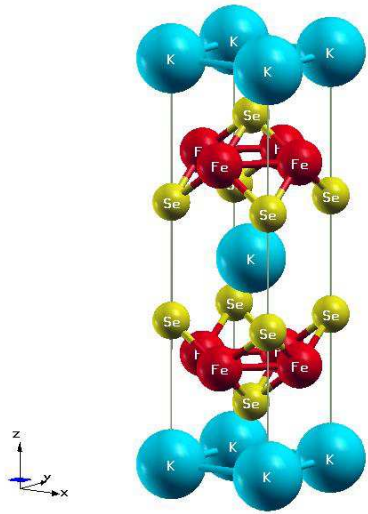}
\caption{Crystal structure of KFe$_{2}$Se$_{2}$ with a tetragonal
unit cell (space group I4/mmm), the axes \emph{X} and \emph{Y} are along the
diagonal Fe-Se direction. The crystal structure parameters are
adopted from Ref. \cite{PRB82-180520}.} \label{fig1}
\end{figure}
%-------------------------------------------------------------------------

In KFe$_{2}$Se$_{2}$, there are formally 6.5 electrons per Fe,
rather than 6 in other iron-based parent materials. Thus it is
regarded as an electron over-doped 11 system.
As a consequence, the hole-like Fermi surface pockets possessed in
other iron-based materials disappear, and only the electron-like
Fermi surface pockets were observed in recent angle-resolved
photoemission spectroscopy (ARPES) experiments
\cite{nmat10-273,PRL106-187001} and predicted in recent band
structure calculations \cite{arXiv1012.5536,arXiv1012.5621}. Thus,
the Fermi surface nesting, which is widely suggested to be the
origin of the striped-antiferromagnetic (AFM) ($\pi$, 0) spin density
wave (SDW) in FeAs-based materials, together with the inter-band
scattering of the Fermi surface, which is widely believed to be
important for the superconducting pairing in FeAs-based
superconductor, is absent in KFe$_{2}$Se$_{2}$ compound.
% ARPES
And the strongly three-dimensional characteristic of the Fermi surface is
also found in KFe$_{2}$Se$_{2}$.

%  LDA -- magnetic properties
Early experiment showed that both TlFe$_{2}$Se$_{2}$
\cite{JSSC63-401} and TlFeS$_{2}$ \cite{JMMM54-57-1497} are AFM
compounds. In a recent experiment, the high-temperature magnetic
susceptibility decreases with the decrease of the temperature,
indicating that the Fe spins interact with each other through the
AFM coupling \cite{arXiv1012.5552}. However, the magnetic
ordering structure is still an debating issue, as addressed in what
follows.
The local density approximation (LDA) calculations suggested that
KFe$_{2}$Se$_{2}$ is a striped AFM order, same to the 1111 and
122 phases of the FeAs-based materials \cite{arXiv1012.5621}, while
some other authors \cite{arXiv1012.5536} thought it to be
bi-collinear AFM with ($\pi$/2, $\pi$/2) wave-vector, similar to
FeTe \cite{PRL102-247001}.
On the other hand, for Fe-deficient materials,
K$_{0.8}$Fe$_{1.6}$Se$_{2}$ {\it etc.}, the block checkerboard AFM
order is observed experimentally, in which its origin is attributed
to the ordered Fe vacancy \cite{arXiv1102.0830,arXiv1102.2882}.
Moreover, the density function theory (DFT) calculations suggested
the TlFe$_{2}$Se$_{2}$ is a checkerboard AFM order
\cite{PRB79-094528}. These discrepant results indicate that further
investigations on the magnetic ordering structure are urgent for
understanding the unique properties in AFe$_{2}$Se$_{2}$ (A=K, Tl, or
Cs) compounds.
In this paper, we have made our efforts to resolve this debate. Based on our LDA
calculated results, we propose a three-dimensional five-orbital
model for KFe$_{2}$Se$_{2}$, and show its dynamical spin
susceptibility diverges at wavevector $\mathbf{Q}$=($\pi$, $\pi$, 0), suggesting
C-type AFM magnetic instability, {\it i.e.} a {\it N$\acute{e}$el}-type AFM
coupling in the $\emph{ab}$-plane and a weak ferromagnetic coupling along the
$\emph{c}$-axis.
This paper is organized as follows: the LDA band structure, Fermi
surface and a five-orbital tight-binding model fitting to the band
structures are presented in {\it Sec. II}; the dynamical spin
susceptibility obtained in the random phase approximation (RPA) is
shown in {\it Sec. III}; the last section is devoted to the remarks
and summary.

\section{Tight-Binding Fitting of the Band Structures}

The parent material KFe$_{2}$Se$_{2}$ has a tetragonal layered
structure with Fe atoms forming a square lattice in the high
temperature phase. The experimental lattice parameters are
$\emph{a}$=$\emph{b}$=3.9136$\AA$ and $\emph{c}$=14.0367$\AA$ at room temperature
\cite{PRB82-180520}.
%
%Because of the tetrahedral coordination of Se, there are two Fe atoms
%per unit cell.
%
There are two Fe atoms per unit cell, where each Fe is tetrahedrally
coordinated by Se.
We have performed a band structure calculation of
KFe$_{2}$Se$_{2}$ using the full-potential linearized augmented
plane-wave plus local orbitals (FP-LAPW+lo) scheme implemented in
the WIEN2K package \cite{wien2k}. The band structure calculations
for other iron-based materials showed that the electronic structure
is sensitive to a small structural distortion. Thus, in order to
compare with the experiments, we adopted the experimental structural
data \cite{PRB82-180520}.
%
% without structural optimization.
%
%And all the band structure calculations throughout this paper are based on the
%experimental structure data \cite{PRB82-180520}.
%
The density of states (DOS), the Fermi surface and the band
structures within the LDA are displayed in Fig. 2, Fig. 3 and Fig. 4,
respectively.
%
%-------------------------------------------------------------------------
\begin{figure}[htbp]\centering
\includegraphics[angle=0, width=0.6 \columnwidth]{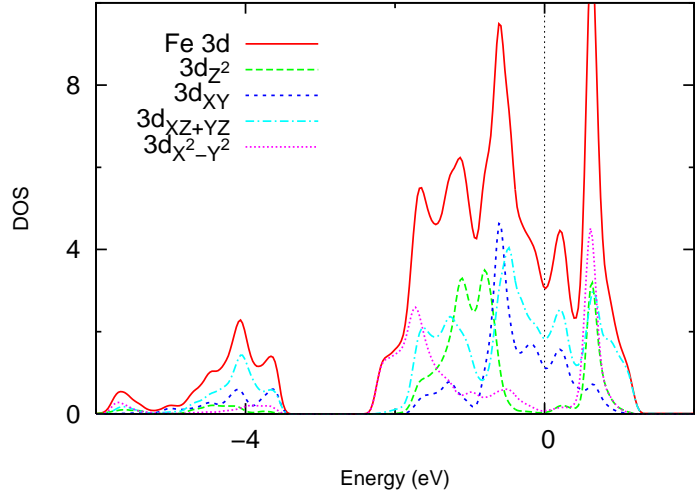}
\caption{The total and partial density of states of Fe obtained by LDA.}
\label{fig2}
\end{figure}
%-------------------------------------------------------------------------
From the DOS in Fig. 2, it is obviously found that this compound is
a metal  in high-temperature phase, and the Fermi surface is mainly
contributed by three orbitals, $XZ$, $YZ$ and $XY$.
%So the corners of Fermi surface occupy these three orbits.
Meanwhile the $\emph{d$_{3Z^{2}-R^{2}}$}$ and $\emph{d$_{X^{2}-Y^{2}}$}$ bands mainly
distribute from $-$2.6 eV to E$_{F}$, with a little contribution to the
Fermi surface, suggesting that the KFe$_{2}$Se$_{2}$ is essentially a
five-band system, similar to other iron pnictides.
%-------------------------------------------------------------------------
\begin{figure}[htbp]\centering
\includegraphics[angle=0, width=0.4 \columnwidth]{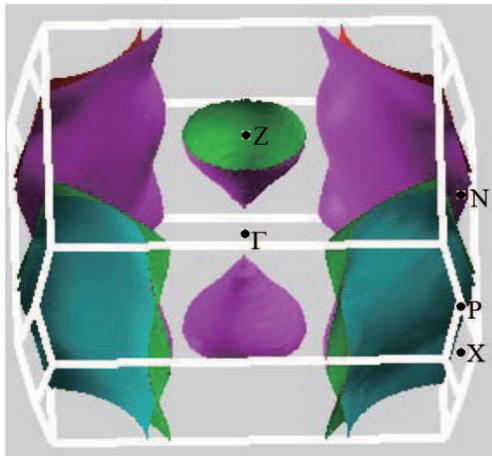}
\caption{Fermi surface obtained within the LDA for parent material
KFe$_{2}$Se$_{2}$ in folded Brillouin zone.} \label{fig3}
\end{figure}
%-------------------------------------------------------------------------
The Fermi surface of KFe$_{2}$Se$_{2}$ is plotted in Fig. 3,
displaying a three-dimensional characteristic with four electron-like Fermi
surface pockets at corner, and two electron-like Fermi surface
pockets along $\emph{$\Gamma$}$-$\emph{Z}$ direction in folded Brillouin zone.
The hole-like Fermi surface pockets at $\emph{$\Gamma$}$
are absent, and only the electron-like Fermi
surface pockets are presented, which is in agreement with the recent
ARPES experiments \cite{nmat10-273,PRL106-187001} and band structure
calculations \cite{arXiv1012.5536,arXiv1012.5621}.

The band structures of the undoped KFe$_{2}$Se$_{2}$ in
high-temperature phase are shown in Fig. 4. To implement our further
study, we fit the band structures with a five-orbital tight-binding
model. Note that the orientation of the coordinate system is chosen
so that Fe-Fe bonds are directed along the $\emph{x}$ and $\emph{y}$
axes, in which $\emph{x}$ and $\emph{y}$ axes are rotated by 45 degrees
from the $\emph{X}$-$\emph{Y}$ axes, where the $\emph{X}$ and
$\emph{Y}$ axes, along the diagonal Fe-Se direction, refer to the original
unit cell. The tight-binding Hamiltonian for the five-orbital model
is described as,
\begin{eqnarray}
   H_{0} &=&
   \sum_{\substack{i,j\\ \alpha,\beta,\sigma}}t_{ij}^{\alpha\beta}
   C_{i\alpha\sigma}^{\dag}C_{j\beta\sigma}-\mu\sum_{i\alpha\sigma}n_{i\alpha\sigma}
\end{eqnarray}
where $C_{i\alpha\sigma}^{\dag}$ creates an electron on site $\emph{i}$ with orbital $\alpha$
and spin $\sigma$, $t_{ij}^{\alpha\beta}$ is the hopping integral between the $\emph{i}$
site with $\alpha$ orbital and the $\emph{j}$ site with $\beta$ orbital, and $\mu$ is
the chemical potential determined by the electron filling.
In the momentum space the Hamiltonian H$_{0}$ is expressed as
\begin{eqnarray}
   H_{0} &=&
   \sum_{\substack{k,\alpha,\beta,\sigma}}(\epsilon_{\alpha}\delta_{\alpha\beta}+
   T^{\alpha\beta}(\textbf{k})) C_{k\alpha\sigma}^{\dag}C_{k\beta\sigma},
\end{eqnarray}
where $T^{\alpha\beta}(\textbf{k})$ is the kinetic energy term, and
$\epsilon_{\alpha}$ denotes the on-site energy of the $\alpha$
orbital. The five-orbital tight-binding fit of the 10
Fe-3d bands obtained by the density functional theory band
structure is displayed in Fig. 4.
%-------------------------------------------------------------------------
\begin{figure}[htbp]\centering
\includegraphics[angle=0, width=0.6 \columnwidth]{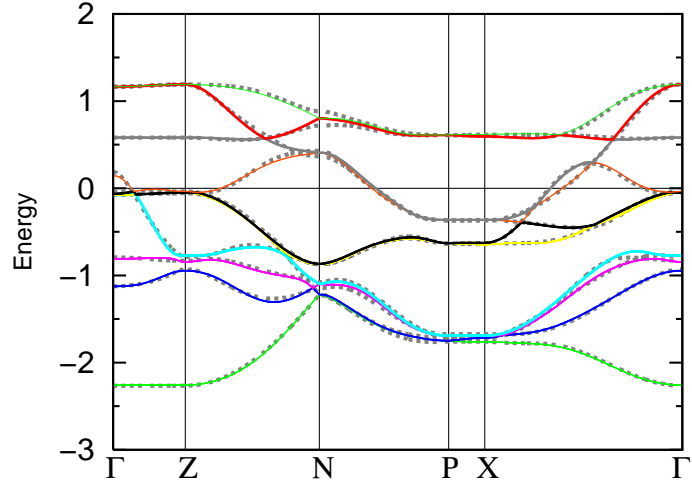}
\caption{The band structures of the Fe-3d orbitals obtained by the
full-potential linearized-augmented-plane-wave (LAPW)
and its five-orbital tight-binding fitting.
The dot lines are the present local-density approximation
results, the solid lines are the fitting results.
The energies are measured from the Fermi energy E$_{F}$=0.409 eV.} \label{fig4}
\end{figure}
%-------------------------------------------------------------------------

The model parameters for the five-orbital tight-binding fitting of
the KFe$_{2}$Se$_{2}$ band structure are listed in the following.
The on-site energies measured from the Fermi energy for the
five orbitals are ($\epsilon_{1},\epsilon_{2},\epsilon_{3},
\epsilon_{4},\epsilon_{5}$)=($-$346.7, $-$346.7, $-$437.4,
$-$176.4, $-$927.9), respectively, in units of meV.
Here orbital indices (1,2,3,4,5) indicate the $\emph{d$_{xz}$}$,
$\emph{d$_{yz}$}$, $\emph{d$_{x^{2}-y^{2}}$(d$_{XY}$)}$, $\emph{d$_{xy}$(d$_{X^{2}-Y^{2}}$)}$,
and $\emph{d$_{3z^{2}-r^{2}}$}$ components, respectively.
Similar to BaFe$_{2}$As$_{2}$ \cite{PRB81-214503},
the hopping integrals along each direction are
\begin{eqnarray}
  T^{11/22} &=& 2t_{x/y}^{11}cos k_{x}+2t_{y/x}^{11}cos k_{y}+4t_{xy}^{11}cos k_{x}cos k_{y} \nonumber \\
  &&\pm 2t_{xx}^{11}(cos 2k_{x}-cos 2k_{y})+4t_{xxy/xyy}^{11}cos 2k_{x}cos k_{y} \nonumber \\
  &&+4t_{xyy/xxy}^{11}cos 2k_{y}cos k_{x}+4t_{xxyy}^{11}cos 2k_{x}cos 2k_{y} \nonumber \\
  &&+4t_{xz}^{11}(cos k_{x}+cos k_{y})cos k_{z}\pm 4t_{xxz}^{11}(cos 2k_{x}-cos 2k_{y})cos(k_{z}), \nonumber
\end{eqnarray}
\begin{eqnarray}
  T^{33}&=& 2t_{x}^{33}(cos k_{x}+cos k_{y})+4t_{xy}^{33}cos k_{x}cos k_{y}+2t_{xx}^{33}(cos 2k_{x}+cos 2k_{y}), \nonumber
\end{eqnarray}
\begin{eqnarray}
  T^{44} &=& 2t_{x}^{44}(cos k_{x}+cos k_{y})+4t_{xy}^{44}cos k_{x}cos k_{y}+2t_{xx}^{44}(cos 2k_{x}+cos 2k_{y}) \nonumber \\
  &&+4t_{xxy}^{44}(cos 2k_{x}cos k_{y}+cos 2k_{y}cos k_{x})+4t_{xxyy}^{44}cos 2k_{x}cos 2k_{y} \nonumber \\
  &&+2t_{z}^{44}cos k_{z}+4t_{xz}^{44}(cos k_{x}+cos k_{y})cos k_{z}+8t_{xyz}^{44}cos k_{x} cos k_{y} cos k_{z}, \nonumber
\end{eqnarray}
\begin{eqnarray}
  T^{55}&=& 2t_{x}^{55}(cos k_{x}+cos k_{y})+4t_{xy}^{55}cos k_{x}cos k_{y}+2t_{xx}^{55}(cos 2k_{x}+cos 2k_{y}) \nonumber \\
  &&+4t_{xxy}^{55}(cos 2k_{x}cos k_{y}+cos 2k_{y}cos k_{x})+4t_{xxyy}^{55}cos 2k_{x}cos 2k_{y} \nonumber \\
  &&+2t_{z}^{55}cos k_{z}+4t_{xz}^{55}(cos k_{x}+cos k_{y})cos k_{z}, \nonumber
\end{eqnarray}
\begin{eqnarray}
  T^{12}&=& 4t_{xy}^{12}sin k_{x}sin k_{y}+4t_{xxy}^{12}(sin 2k_{x} sin k_{y}+sin 2k_{y} sin k_{x}) \nonumber \\
  &&+4t_{xxyy}^{12}sin 2k_{x}sin 2k_{y}+8t_{xyz}^{12}sin k_{x} sin k_{y} cos k_{z}, \nonumber
\end{eqnarray}
\begin{eqnarray}
  T^{13/23}&=& 2it_{x}^{13}sin k_{y/x}+4it_{xy}^{13}cos k_{x/y} sin k_{y/x} \nonumber \\
  &&-4it_{xxy}^{13}(sin 2k_{y/x} cos k_{x/y}-cos 2k_{x/y} sin k_{y/x}), \nonumber
\end{eqnarray}
\begin{eqnarray}
  T^{14/24}&=& \pm 2it_{x}^{13}sin k_{x/y}\pm 4it_{xy}^{14}cos k_{y/x} sin k_{x/y} \nonumber \\
  &&\pm 4it_{xxy}^{14}cos k_{y/x} sin 2k_{x/y}\pm 4it_{xz}^{14}sin k_{x/y} cos k_{z} \nonumber \\
  &&-4t_{xz}^{24}sin k_{x/y} sin k_{z}\pm 8it_{xyz}^{14}cos k_{y/x} sin k_{x/y}cos k_{z} \nonumber \\
  &&\pm 8it_{xxyz}^{14}sin 2k_{x/y} cos k_{y/x} cos k_{z}-8t_{xxyz}^{24}sin 2k_{x/y} cos k_{y/x} sin k_{z}, \nonumber
\end{eqnarray}
\begin{eqnarray}
  T^{15/25}&=& \pm 2it_{x}^{15}sin k_{y/x}\mp 4it_{xy}^{15}cos k_{x/y} sin k_{y/x} \nonumber \\
  &&\mp 8it_{xyz}^{15}cos k_{x/y} sin k_{y/x} cos k_{z}, \nonumber
\end{eqnarray}
\begin{eqnarray}
  T^{34}&=& 4t_{xxy}^{34}(sin 2k_{y} sin k_{x}-sin 2k_{x} sin k_{y}), \nonumber
 \end{eqnarray}
\begin{eqnarray}
  T^{35}&=& 2t_{x}^{35}(cos k_{x}-cos k_{y})+4t_{xxy}^{35}(cos 2k_{x} cos k_{y}-cos 2k_{y} cos k_{x}). \nonumber
\end{eqnarray}
\begin{eqnarray}
  T^{45}&=& 4t_{xy}^{45}sin k_{x} sin k_{y}+4t_{xxyy}^{45}sin 2k_{x} sin 2k_{y} \nonumber \\
  &&+2it_{z}^{45}sin k_{z}+4it_{xz}^{45}(cos k_{x}+cos k_{y})sin k_{z}, \nonumber
 \end{eqnarray}

%Similar to Ref. \cite{PRB81-214503},
The intra-orbital and inter-orbital hopping parameters up to fifth neighbors of the five-orbital
model for the fit of the band structure are shown in Table I.
%----------------------------------------------------------------------
\begin{table}[htbp]
\begin{center}
\caption{The intra-orbital t$_{i}^{\alpha\alpha}$ and
inter-orbital t$_{i}^{\alpha\beta}$ hopping parameters up to fifth neighbors
of the five orbital tight-binding model through fitting
the band structures. All the parameters are in units of meV.}
 \vspace{0.2cm}
\begin{tabular}{lcccccccccccccc}
\hline \hline
&t$_{i}^{\alpha\alpha}$ &i=x     &i=y     &i=xy   &i=xx    &i=xxy   &i=xyy  &i=xxyy  &i=z    &i=xz    &i=xxz  &i=xyz \\
\hline
&$\alpha$=1             &$-$15.6 &$-$236.7&198.2  &10.5    &$-$42.2 &13.1   &27.9    &       &$-$1.7    &12.2   &    \\
\hline
&$\alpha$=3             &355.0   &        &$-$77.4&$-$22.7 &        &       &        &       &          &       &  \\
\hline
&$\alpha$=4             &$-$3.4  &        &54.5   &$-$28.0 &$-$17.0 &       &$-$32.3 &66.0   &30.1      &       &15.3  \\
\hline
&$\alpha$=5             &66.1    &        &       &6.5     &$-$17.8 &       &$-$2.7  &18.0   &$-$6.8    &       & \\
\hline \hline
&t$_{i}^{\alpha\beta}$  &i=x     &i=xy    &i=xxy   &i=xxyy    &i=z    &i=xz    &i=xyz    &i=xxyz \\
\hline
&$\alpha\beta$=12       &        &41.3    &0.7     &21.3      &       &        &45.9     &   \\
\hline
&$\alpha\beta$=13       &$-$302.7&122.5   &22.6    &          &       &        &         &    \\
\hline
&$\alpha\beta$=14       &$-$213.2&$-$26.2 &1.7     &          &       &$-$11.3 &4.9      &$-$10.6 \\
\hline
&$\alpha\beta$=15       &$-$87.3 &$-$103.3&        &          &       &        &7.8      &  \\
\hline
&$\alpha\beta$=24       &        &        &        &          &       &$-$24.5 &         &13.9    \\
\hline
&$\alpha\beta$=34       &        &        &1.7     &          &       &        &         &     \\
\hline
&$\alpha\beta$=35       &$-$293.9&        &$-$6.9  &          &       &        &         &    \\
\hline
&$\alpha\beta$=45       &        &68.0    &        &$-$32.7   &$-$65.6&7.7     &         &    \\
\hline \hline
\end{tabular}
\end{center}
\end{table}
%------------------------------------------------------------------------
%
%

\section{Dynamical Spin susceptibility}

In this section we study the dynamical spin susceptibility
of the five-orbital tight-binding model for KFe$_{2}$Se$_{2}$.
The dynamical magnetic susceptibility is calculated
for both the non-interaction case and the electron-electron
interaction case within the RPA.

%We firstly study the bare spin susceptibility,
%then consider the electron-electron interaction
%on the spin fluctuations within random phase approximation (RPA).

The orbital-dependent dynamical spin susceptibility is
given by \cite{PRB75-224509,NJP11-025016}
\begin{eqnarray}
  \chi_{\alpha\gamma}(\mathbf{q}, i\omega) &=& \int_{0}^{\beta}d\tau e^{i\omega\tau}<T_{\tau}\mathbf{S}_{\alpha}(\mathbf{q},\tau)\cdot\mathbf{S}_{\gamma}(-\mathbf{q},0)>
\end{eqnarray}
where $\alpha$ and $\gamma$ label the orbital indices, and the spin operator $\mathbf{S}_{\alpha}(\mathbf{q})$=$\frac{1}{2}\sum_{\substack{k,mn}}C_{\alpha m}^{\dag}(\mathbf{k}+\mathbf{q})\mathbf{\sigma}_{mn}C_{\alpha n}(\mathbf{k})$ with spin indices m,n, and $\beta$=1/$k_{B}T$.
The physical dynamical spin susceptibility reads $\chi(\mathbf{q},i\omega)=\frac{1}{2}\sum_{\substack{\alpha\gamma}}\chi_{\alpha\gamma}(\mathbf{q},i\omega)$, with
\begin{eqnarray}
  \chi_{\alpha\gamma}(\mathbf{q}, i\omega) &=& -\frac{1}{N}\sum_{\substack{k,\mu\nu}}\frac{b_{\mu}^{\alpha}(\mathbf{k})b_{\mu}^{\gamma *}(\mathbf{k})b_{\nu}^{\gamma}(\mathbf{k}+\mathbf{q})b_{\nu}^{\alpha *}(\mathbf{k}+\mathbf{q})}{i\omega+E_{\nu}(\mathbf{k}+\mathbf{q})
  -E_{\mu}(\mathbf{k})}[f(E_{\nu}(\mathbf{k}+\mathbf{q}))-f(E_{\mu}(\mathbf{k}))].
\end{eqnarray}
Here, the matrix elements of the eigenvector
$b_{\mu}^{\alpha}(\mathbf{k})=<\alpha|\mu \mathbf{k}>$ with indices connecting $\alpha$
orbital and $\mu$ band, are determined by the diagonalization of
the tight-binding model Hamiltonian Eq. (2).

In the presence of the Coulomb interaction, in addition
to the kinetic term in Eq. (2), the electronic
interaction part of the multiorbital Hamiltonian reads,
\begin{eqnarray}
  H_{I} &=& U\sum_{\substack{i, \alpha}}n_{i\alpha\uparrow}n_{i\alpha\downarrow}
  +U^{'}\sum_{\substack{i\\ \alpha\ne\beta}}n_{i\alpha\uparrow}n_{i\beta\downarrow}
  +(U^{'}-J_{H})\sum_{\substack{i,\sigma\\ \alpha<\beta}}n_{i\alpha\sigma}n_{i\beta\sigma}
\nonumber\\
  &&-J_{H}\sum_{\substack{i\\ \alpha\ne\beta}}
  C_{i\alpha\uparrow}^{\dag}C_{i\alpha\downarrow}C_{i\beta\downarrow}^{\dag}C_{i\beta\uparrow}
  +J_{H}\sum_{\substack{i\\ \alpha\ne\beta}}
  C_{i\alpha\uparrow}^{\dag}C_{i\alpha\downarrow}^{\dag}C_{i\beta\downarrow}C_{i\beta\uparrow}
\end{eqnarray}
where U(U$^{'}$) denotes the intra-(inter-)orbital Coulomb
repulsion interaction and J$_{H}$ the Hund's
rule coupling. Considering the symmetry of the system,
we adopt U$^{'}$=U-2J$_{H}$.
The RPA dynamical magnetic susceptibility is described as
\begin{eqnarray}
  \chi^{RPA}(\mathbf{q}, i\omega) &=& \chi_{0}(\mathbf{q},i\omega)[\mathbb{I}-\Gamma\chi_{0}(\mathbf{q},i\omega)]^{-1}
\end{eqnarray}
where $\chi_{0}$ is the bare spin susceptibility defined in Eq. (4),
and the nonzero components of the matrices $\Gamma$
are given as $(\Gamma)_{\alpha\alpha}=U$,
and $(\Gamma)_{\alpha\neq\gamma}=J_{H}$.
Note that the interorbital interaction
U$^{'}$ doesn't contribute to the RPA spin
susceptibility \cite{NJP11-025016}.
%
%-------------------------------------------------------------------------
\begin{figure}[htbp]\centering
\includegraphics[angle=0, width=0.8 \columnwidth]{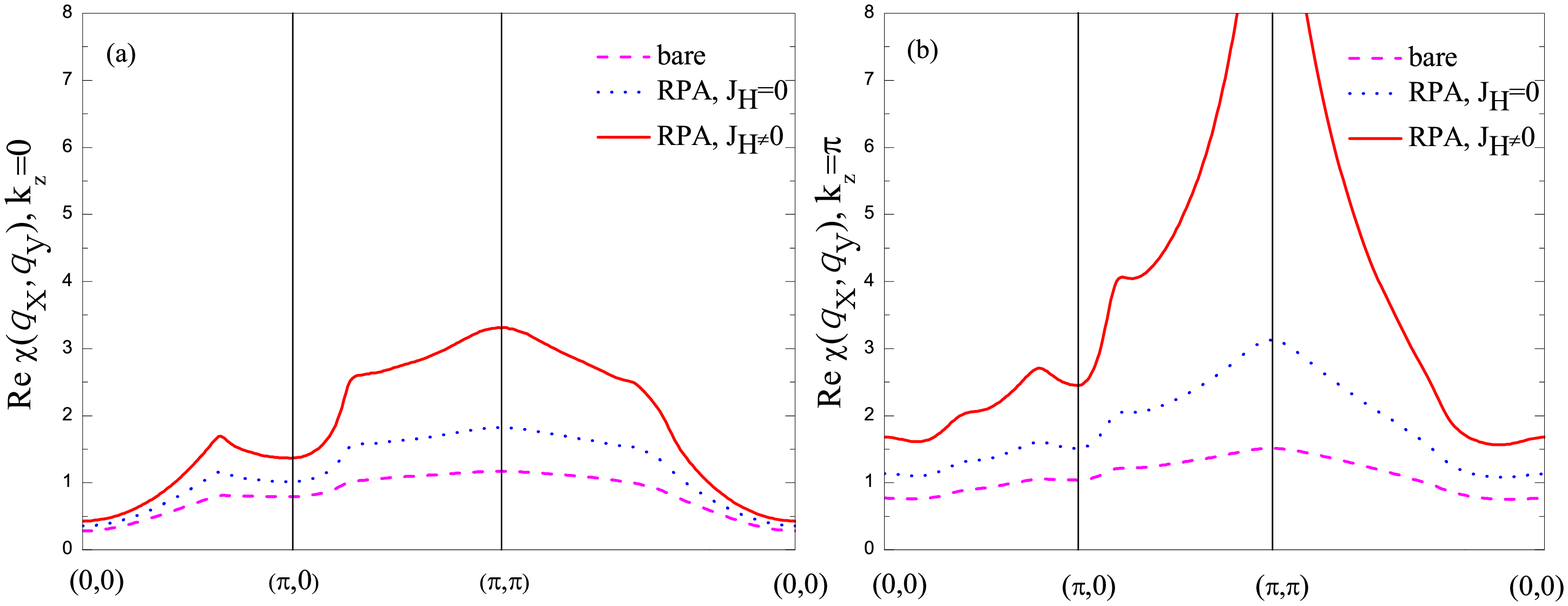}
\caption{Real part of the bare and RPA
enhanced susceptibility along the main symmetry directions of the
Fe/cell Brillouin zone for k$_{z}$=0 (a), $\pi$ (b) in the
two-dimensional case (k$_{z}$ is fixed and q$_{z}$=0).
Theoretical parameters U=0.6 eV, J$_{H}$=0, or 0.25U ($\neq$0),
and $\beta$=0.02.} \label{fig5}
\end{figure}
%-------------------------------------------------------------------------

To investigate the effect of the k$_{z}$-dependence of $\chi(\mathbf{q},
i\omega)$, we study the two-dimensional case with fixed k$_{z}$ and
q$_{z}$=0, and the three-dimensional case with integrating over
k$_{z}$ and fixed q$_{z}$, respectively.
Fig. 5 plots the q-dependence of the real part of the bare and RPA
dynamical magnetic susceptibilities in the two-dimensional case for
$k_{z}$=0 and $\pi$ along the main symmetry directions of the
Fe/cell Brillouin zone, respectively.
It is found that the bare spin susceptibility is nearly a
plateau-like  structure, which is almost equivalent both for
$k_{z}$=0 and for $k_{z}$=$\pi$, without obvious peak.
As a comparison, the RPA dynamical spin susceptibility is
considerably enhanced due to the presence of the Coulomb
interaction, and a small peak appears at $\emph{Q}$=($\pi$, $\pi$).
Furthermore, in the presence of both Coulomb interaction and finite
Hund's rule coupling, J$_{H}$$\ne$0, the ($\pi$, $\pi$)
susceptibility peak is greatly enhanced.
Moreover, the RPA spin susceptibility of the $k_{z}$=$\pi$ cut is
obviously larger than that of the $k_{z}$=0 cut.
In addition to a small peak along the (0, 0)-($\pi$, 0) line, the
largest peak appears at $\mathbf{Q}$=($\pi$, $\pi$), indicating the magnetic
instability of the $\it{N\acute{e}el}$-type AFM order rather than
that of the bi-collinear AFM order with $\mathbf{Q}$=($\pi$/2, $\pi$/2), or of
the striped AFM order with $\mathbf{Q}$=($\pi$, 0).
Similar behaviors are also seen in Fig. 6, which displays the
contour of the real part of the RPA dynamical magnetic
susceptibility for fixed value $k_{z}$=0 (see (a) and (c)),
and $\pi$ (see (b) and (d)) in the q$_{x}$-q$_{y}$ plane
in the two-dimensional case.
%
%-------------------------------------------------------------------------
\begin{figure}[htbp]\centering
\includegraphics[angle=0, width=0.6 \columnwidth]{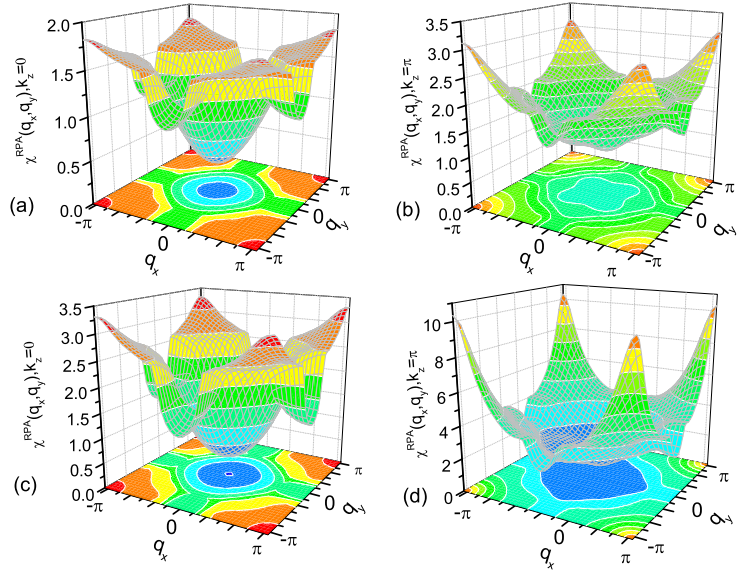}
\caption{Contour of the real part of the RPA
dynamical magnetic susceptibility of k$_{z}$=0 ((a) and (c)),
and $\pi$ ((b) and (d)) in the
two-dimensional case (k$_{z}$ is fixed and q$_{z}$=0).
The theoretical parameters: Coulomb
interaction U=0.6 eV, J$_{H}$=0 ((a) and (b)),
and 0.25U ((c) and (d)), and $\beta$=0.02.}
\label{fig6}
\end{figure}
%-------------------------------------------------------------------------
%

Considering the three-dimensional characteristic of Fermi surface in
KFe$_{2}$Se$_{2}$, as seen in Fig. 3, we also present the k$_{z}$
evolution of the magnetic susceptibility.
Similar to the two-dimensional case, the three-dimensional bare and
RPA dynamical susceptibilities are calculated for fixed q$_{z}$
values, q$_{z}$=0 or $\pi$.
As expected, the peaks of the dynamical magnetic susceptibility are
suppressed in the three-dimensional case, in comparison with the
two-dimensional result. To clearly show the magnetic instability
in the three-dimensional case, we displays the susceptibilities at
U= 0.7 eV, as shown in Fig. 7. We find that the peak grows up and
the divergent tendency of $\chi$($\mathbf{q}$) becomes strong with
the increase of the electronic correlations. In realistic compound
K$_{x}$Fe$_{2}$Se$_{2}$, the Coulomb interaction U is obviously
larger than 0.7 eV, implying that the divergency of the
$\chi$($\mathbf{q}$) is much stronger than the present situation
shown in Fig. 7. Thus the strong divergency in the dynamical
magnetic susceptibility suggests the magnetic instability or spin
ordering in K$_{x}$Fe$_{2}$Se$_{2}$.

%-------------------------------------------------------------------------
\begin{figure}[htbp]\centering
\includegraphics[angle=0, width=0.8 \columnwidth]{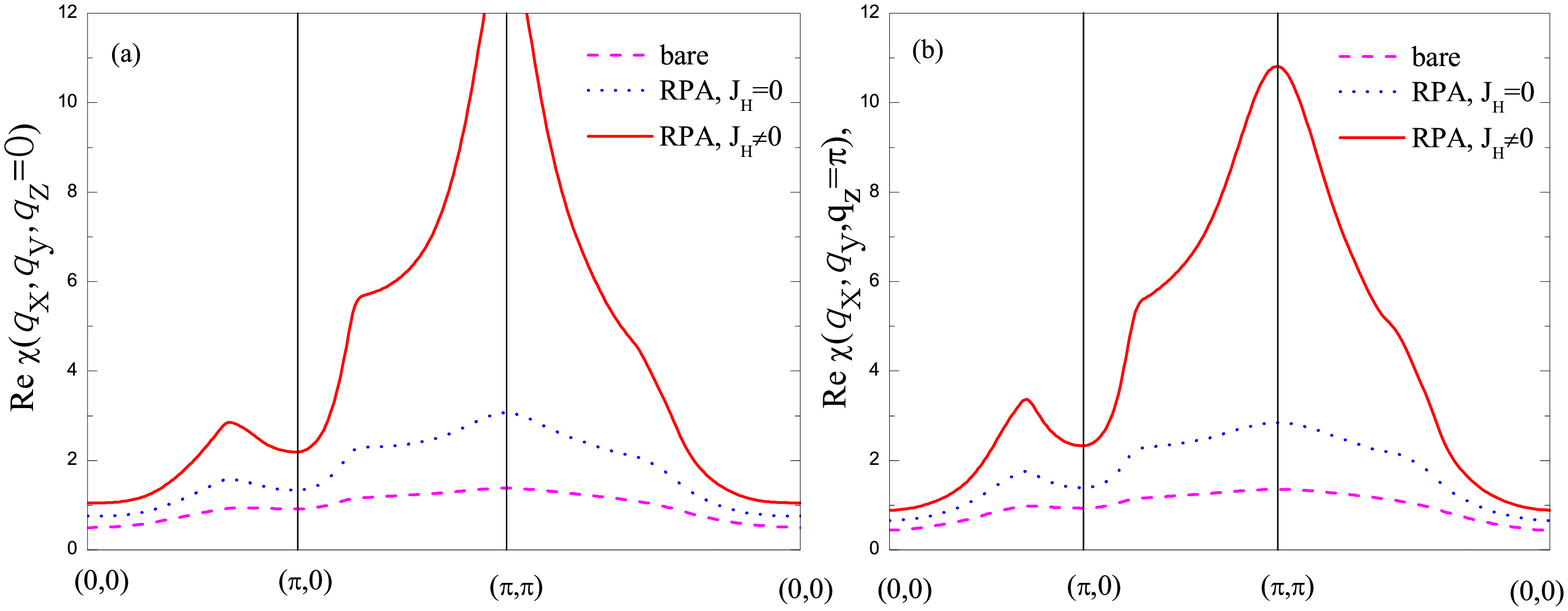}
\caption{Real part of the bare and RPA
dynamical magnetic susceptibility along the main symmetry directions of the
Fe/cell Brillouin zone for q$_{z}$=0 (a), $\pi$ (b) in the
three-dimensional case. We adopt U=0.7 eV,
J$_{H}$=0, 0.25U ($\neq$0), and $\beta$=0.02.}
\label{fig7}
\end{figure}
%-------------------------------------------------------------------------

The dynamical susceptibility in Fig. 7 shows a small peak in the (0,
0)-($\pi$, 0) line and a large peak around at ($\pi$, $\pi$),
similar to the two-dimensional case.
The sharp peak in the spin susceptibility at q$_{z}$=0 indicates a
($\pi$, $\pi$, 0) three-dimensional magnetic structure with
ferromagnetic coupling between the nearest-neighbor FeSe layers in
KFe$_{2}$Se$_{2}$, {\it i.e.} it shows a ferromagnetic coupling
along the $\emph{c}$-axis and an AFM coupling in the $\emph{ab}$-plane (C-type AFM).
Notice that the interlayer coupling is weak due to the long distance
between the nearest-neighbor Fe layers along the $\emph{c}$-axis. Other LDA
calculations \cite{arXiv1012.5536} for both CsFe$_{2}$Se$_{2}$ and
KFe$_{2}$Se$_{2}$ also suggested a ferromagnetic coupling along $\emph{c}$
axis, in agreement with our result.
%%
%
% As a contrast, Experiment on TlFe$_{2}$Se$_{2}$ demonstrate an AFM
% interlayers \cite{PRB79-094528}.
% In the Fe-vacancy ordered phase, {\it e.g.}
% K$_{0.8}$Fe$_{1.6}$Se$_{2}$, the AFM between the nearest neighboring
% layers FeSe is also observed in the neutron scattering experiments
% \cite{arXiv1102.0830, arXiv1102.2882}. And LDA calculations for
% AFe$_{1.5}$Se$_{2}$ (A=K, Rb, or Cs) \cite{PRL106-087005} also found
% that it favors AFM interlayer coupling.
%
%
% The discrepancies among these materials suggest the magnetism is
% sensitive to the lattice distortion or the electronic correlation.
%
However, the structures of the spin susceptibility of q$_{z}$=$0$ and
$\pi$ remain qualitatively the same, as shown in Fig. 8.
%-------------------------------------------------------------------------
\begin{figure}[htbp]\centering
\includegraphics[angle=0, width=0.6 \columnwidth]{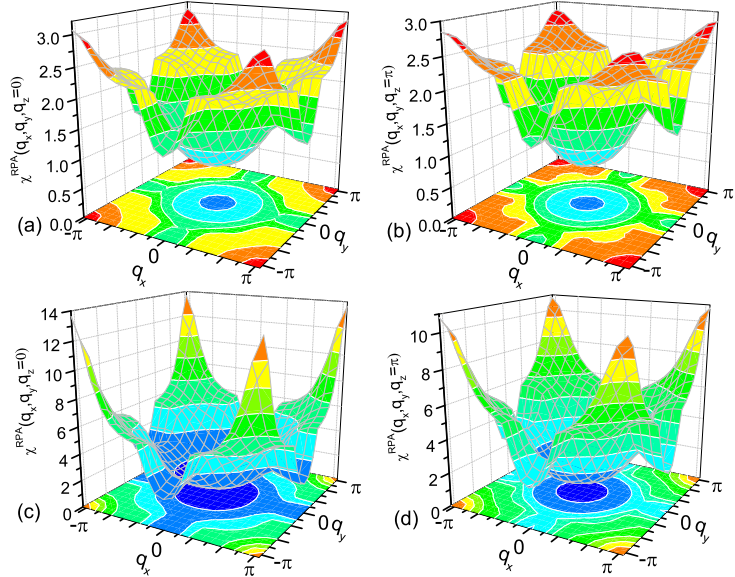}
\caption{Contour of the real part of the RPA
dynamical magnetic susceptibility for q$_{z}$=0 ((a) and (c)),
and $\pi$ ((b) and (d)) in the
three-dimensional case. The Coulomb
interaction U=0.7 eV, J$_{H}$=0 (a) and (b),
0.25U (c) and (d), and $\beta$=0.02 are adopted.}
\label{fig8}
\end{figure}
%-------------------------------------------------------------------------

On the contrary, the first-principles electronic structure
calculations suggested the striped AFM in KFe$_{2}$Se$_{2}$
\cite{arXiv1012.5621}, or the bi-collinear AFM ordering
%
% resulting from the interplay among the nearest-, the next-nearest, and the
%next-next-nearest-neighbor superexchange interactions mediated by Se
%4p-orbitals
%
in AFe$_{2}$Se$_{2}$ (A=K, Tl, or Cs) \cite{arXiv1012.5536}. Though
Zhang and Singh's {\it N$\acute{e}$el}-type AFM configuration in
TlFe$_{2}$Se$_{2}$ \cite{PRB79-094528} agrees with our results in
the $\emph{ab}$-plane, the magnetic coupling along the $\emph{c}$-axis
is different.
The origin of the difference in these {\it ab initio} results is
unknown. In our study, we deal with the electronic correlation in
the framework of the RPA. Considering the fact that the FeSe-based
materials with large magnetic moments may be intermediate or strong
correlated systems, a proper treatment on the electronic correlation
in these FeSe-based compounds may be important for understanding its
magnetic ground state.

On the other hand, it is found in the recent experiments that the
ordered Fe vacancies in K$_{0.8}$Fe$_{1.6}$Se$_{2}$ lead to a
block checkerboard AFM order in the square Fe layers with a large
magnetic moment 3.31 $\mu_{B}$ and T$_{N}$ $\sim$ 559 K
\cite{arXiv1102.0830}, then an Fe vacancy order-disorder transition
occurs at T$_{S}$ $\sim$ 578 K. In addition, it is also shown
that the AFM order is reduced \cite{arXiv1012.5236} with the
increasing Fe content. The block checkerboard AFM order observed in
the experiments of K$_{0.8}$Fe$_{1.6}$Se$_{2}$ is confirmed in
recent LDA calculations \cite{arXiv1102.2215}.
Meanwhile, AFe$_{1.5}$Se$_{2}$ (A=K, Tl, Rb, or Cs) displays a
A-collinear AFM by the LDA calculations \cite{PRL106-087005}.
Obviously, our prediction on the C-type AFM magnetic ordering in
KFe$_{2}$Se$_{2}$ is different from these experimental observations,
showing that Fe vacancy with the lattice dynamics plays a key role
in these Fe-deficient materials.
%
% both the experimental and the theoretical LDA results shown that the
% magnetism in these systems is sensitive to the lattice distortion,
% which suggests the magnetic phase transition is associated with the
% structural transition.
%
% We can conclude that the magnetism may be closely associated with
% the lattice distortion, thus the orbital ordering scenario should be
% involved in the future work to unveil the conflicting results.

\section{Summary}

In summary, we have presented a five-orbital tight-binding model for
newly found KFe$_{2}$Se$_{2}$, the energy dispersion and Fermi
surface are also given. In addition, we also investigated the spin
fluctuation through the multi-orbital dynamical spin susceptibility
within the random phase approximation. The results demonstrate a
divergent peak appears at the wavevector $\mathbf{Q}$=($\pi$, $\pi$, 0),
indicating a C-type antiferromagnetic ordering in the parent
compound KFe$_{2}$Se$_{2}$. Future experiment for parent materials
KFe$_{2}$Se$_{2}$ are expected to confirm this kind of magnetic
ordering. Further theoretical studies on the dynamical magnetic
susceptibility for the Fe-deficient compounds are also deserved.

\acknowledgements

The author (D.Y.) gratefully acknowledge the help by Guo-Ren Zhang
and Yan-Ling Li. This work was supported by the Natural Science
Foundation of China (NSFC) No. 11047154, 11074257, and of Anhui
Province No. 11040606Q56, and the Knowledge Innovation Program of
the Chinese Academy of Sciences. Numerical calculations were
performed at the Center for Computational Science of CASHIPS.

\bibliography{apsrev4-1}% Produces the bibliography via BibTeX.

\end{document}